\documentclass[a4paper,11pt]{article}
\pdfoutput=1 

\usepackage{jinstpub} 

\title{\boldmath Detailed Measurements of Shower Properties in a High Granularity Digital Electromagnetic Calorimeter}

\author{Naomi van der Kolk}
\affiliation{Nikhef,\\Amsterdam, the Netherlands}

\emailAdd{naomi.van.der.kolk@cern.ch}

\abstract{The MAPS prototype of the proposed ALICE Forward Calorimeter (FoCal) is the highest granularity electromagnetic calorimeter, with 39 million pixels of $30 \times 30 \mu$m$^{2}$. Particle showers can be studied with unprecedented detail with this prototype. Electromagnetic showers at energies between 2\,GeV and 244\,GeV have been studied and compared to Geant4 simulations. Simulation models can be tested in more detail than ever before and the differences observed between FoCal data and Geant4 simulations illustrates that improvements in electromagnetic models are still possible.}


\collaboration[c]{on behalf of the ALICE FoCal collaboration}

\proceeding{CHEF 2017\\
  2-6 October 2017\\
  Lyon, France}

\begin{document}
\maketitle
\flushbottom

\section{Introduction}
\label{sec:intro}

In light of the upgrade program of the ALICE detector~\cite{upgrade} a calorimeter at forward rapidities (FoCal) is being considered.
This detector would measure photons, electrons, positrons and jets for rapidities $\eta > 3$ offering a wealth of physics possibilities.
Its main focus is on measurements related to the structure of nucleons and nuclei at very low Bjorken-x and possible effects of gluon saturation~\cite{CGC}.
FoCal consists of an electromagnetic calorimeter, that would be most likely positioned at a distance from the IP of $\textrm{z} \approx 7$ m covering $3.2 < \eta < 5.3$,
backed by a hadronic calorimeter.
The electromagnetic calorimeter must be able to discriminate decay photons from direct photons at very high energy, which, in this region of phase space, requires extremely high granularity.
The design option for the electromagnetic calorimeter currently under study is a silicon-tungsten (SiW) sandwich construction.
It consists of 20 layers of a 3.5 mm tungsten plates ($\approx 1 X_{0}$) interleaved with active layers with Si sensors.
The active layers use two different sensor technologies: low granularity layers (LGL), which consist of sensors with 1 $\textrm{cm}^{2}$ pads summed longitudinally in segments and equipped with analog readout, and high granularity layers (HGL) based on CMOS monolithic active pixel sensors (MAPS).
The MAPS will have a pixel size of about $30 \times 30 \, \mu \textrm{m}^{2}$ with internal binary readout.
The HGL are crucial for $\gamma - \pi^{0}$ discrimination.

\begin{figure}[h]
\centering 
\includegraphics[width=.5\textwidth]{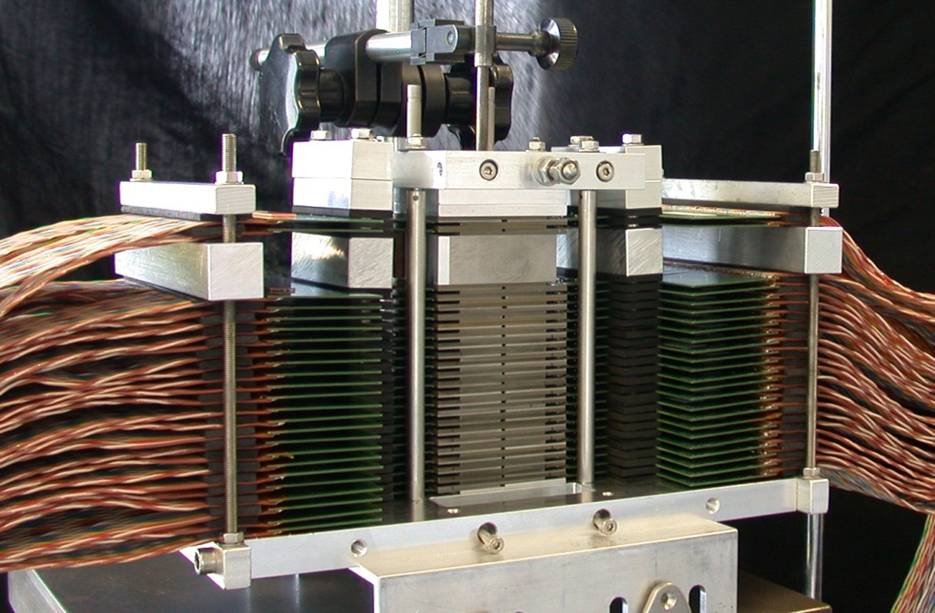}
\caption{\label{fig:focal} A photo of the FoCal prototype.}
\end{figure}

As part of the R\&D program for FoCal a full MAPS calorimeter prototype has been designed and constructed to validate the concept of a calorimeter using high-granularity CMOS pixel sensors with digital readout.
The prototype has a quadratic cross section of $4 \times 4 \, \textrm{cm}^{2}$ and a total depth of $\approx 10 \, \textrm{cm}$.
It consists of 24 layers of 3.3 mm thick tungsten plates and about 1 mm of sensor+infrastructure material.
This highly compact design leads to effective detector parameters of $X_{0} \approx 4$ mm and R$_{\textrm{M}} \approx 11$ mm.
The sensor used is the MIMOSA23 chip~\cite{mimosa}.
The active area of the MIMOSA23 consists of $640 \times 640$ pixels of a surface of $30 \times 30 \, \mu \textrm{m}^{2}$ each.
Every active layer uses $2 \times 2$ chips, providing about 40 million pixels in total.
Every chip is read out in a cycle of 640 $\mu$s by a so-called rolling shutter.
All chips are connected to a cooling system via the tungsten plates.
The prototype setup is complemented with a set of small scintillators for particle triggers.
Figure~\ref{fig:focal} shows a photo of the FoCal prototype.
It, and its performance, is described in detail in~\cite{prototype}.
Because of its extremely high granulairity the FoCal prototype offers an ideal opportunity for testing the models for electromagnetic and hadronic shower development that are being used in MC simulations.

\section{Calibration}
\label{sec:calibration}
\begin{figure}[h]
\centering 
\includegraphics[width=.7\textwidth]{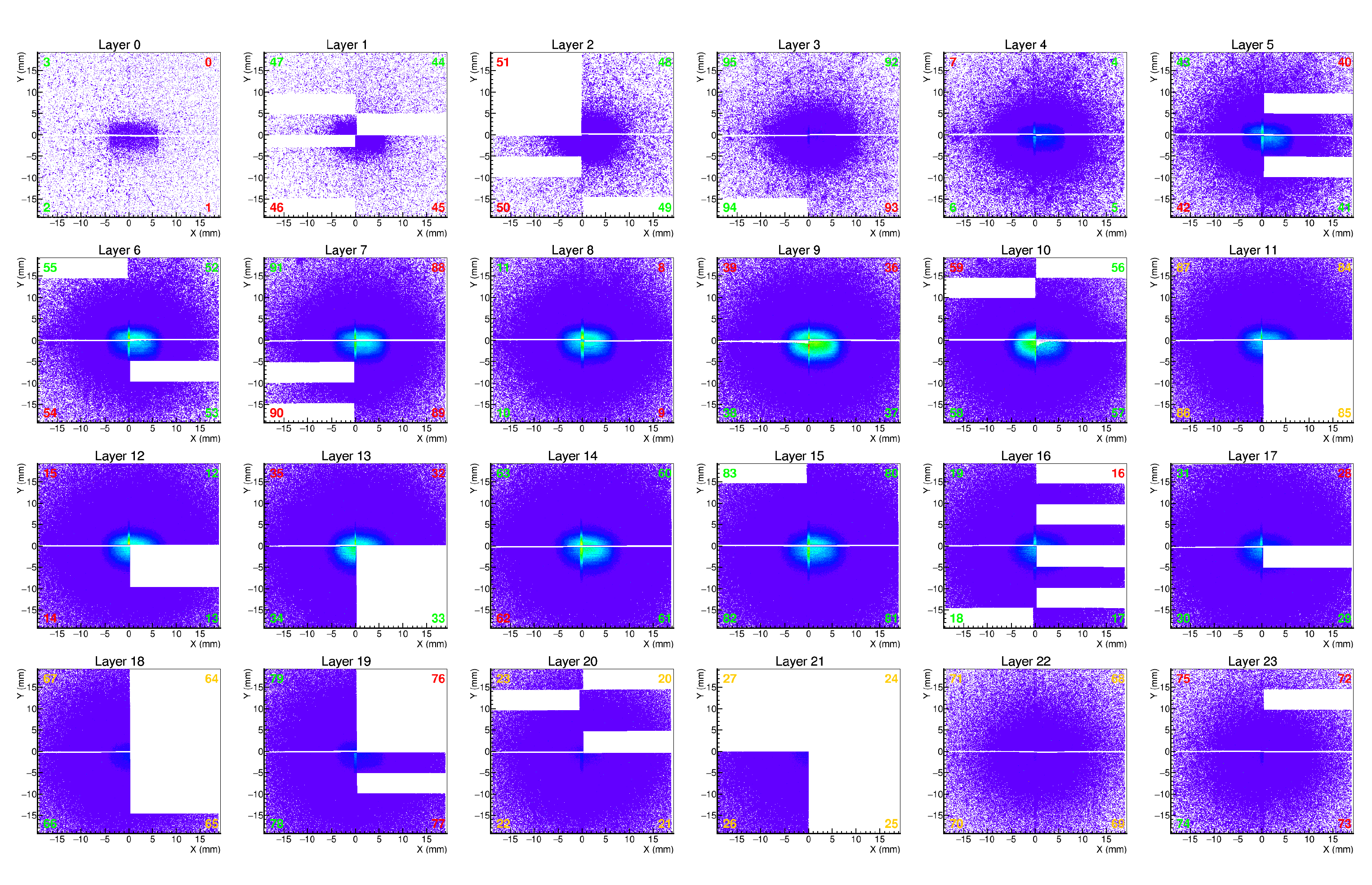}
\caption{\label{fig:hitmap} A hitmap of all 24 of the FoCal prototype layers. One can clearly seen that there are some inefficient areas.}
\end{figure}

\begin{figure}[h]
\centering 
\includegraphics[width=.5\textwidth]{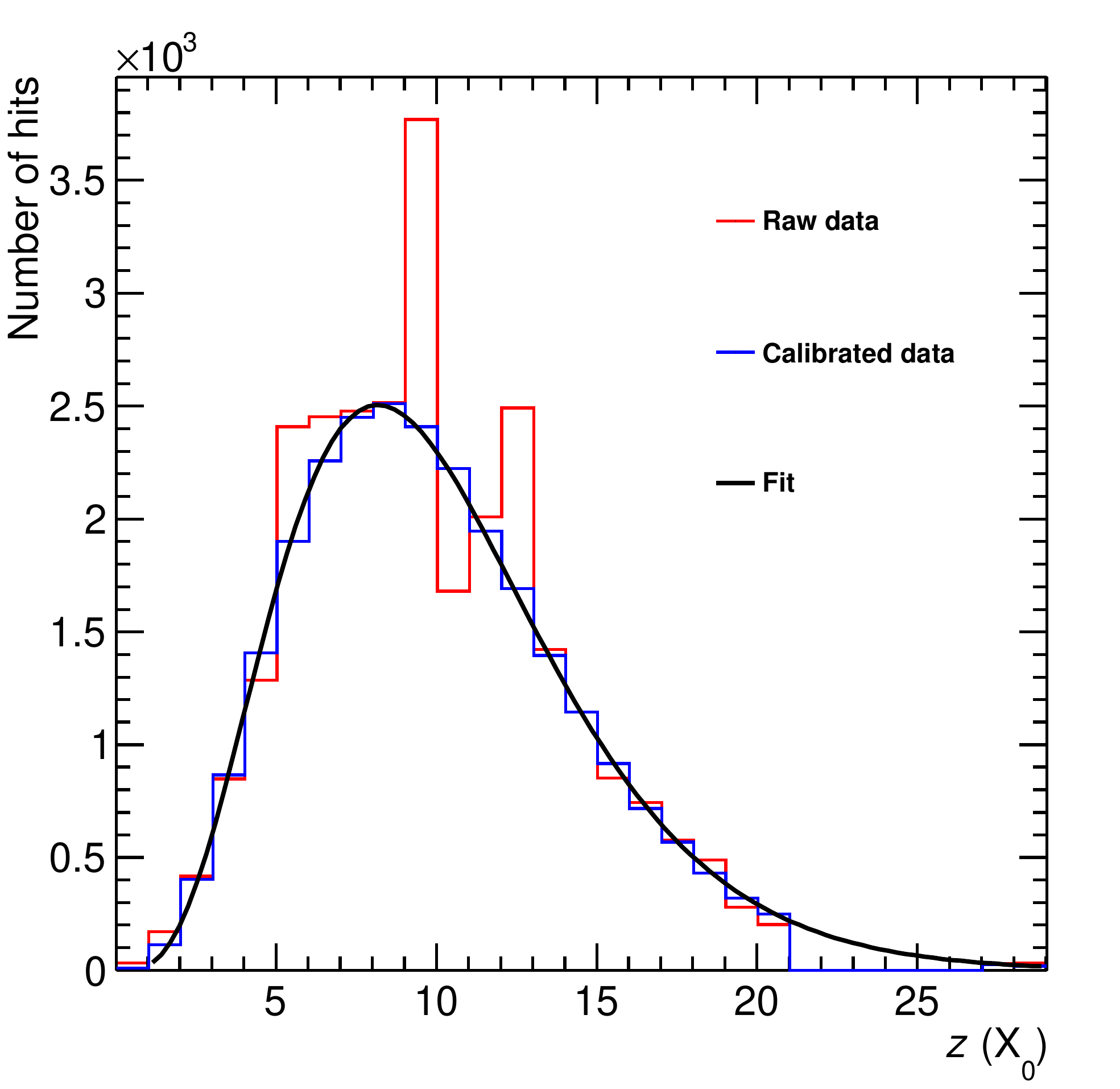}
\caption{\label{fig:nhits} The total number of hits as a function of depth, z($X_{0}$), before and after the calibration procedure.}
\end{figure}

Test beam measurements have been performed at DESY with positron beams (2, 3, 4 and 5.4\,GeV), and with mixed beams at the CERN SPS (positron and $\pi^{+}$ at 30, 50, 100\,GeV, 
and electron and $\pi^{-}$ at 244\,GeV).
The prototype has in addition been exposed to cosmic muons for an extended period of time. 
As can be seen in Fig.~\ref{fig:hitmap} a relatively large portion of the detector, 17\%, was not of sufficient quality to be used in the data taking.
In order to fully reconstruct recorded events a correction method for these unused areas is applied.
The calibration of the prototype also follows from this procedure.
First the shower centre is determined and then the density of hits is calculated for thin rings around this centre.
This is done separately for each of the 4 sensors in one layer.
Then the response of the 4 sensors are equalised, based on the best sensor in that layer.
Unusable areas are compensated for by assuming a symmetrical hit density in electromagnetic showers.
The response of the different layers is equalised assuming a Gamma function in the longitudinal direction;
\begin{equation*}
\label{eq:gamma}
N(t) = N_{0}\, b \, \frac{(bt)^{a-1}e^{bt}}{\Lambda(a)}\, (t = \frac{x}{X_{0}})\,.
\end{equation*}
The effect of the calibration is illustrated in Fig.~\ref{fig:nhits}, where the raw and the calibrated number of hits is shown.
All reconstructed event properties are subsequently based on the calibrated hit density as a function of radius and layer.

The prototype has been modelled in a MC simulation using GEANT4~\cite{G4}.
The full geometry is taken into account, including the detector inefficiencies and noise, which is on the level of $10^{-5}$ per pixel.
A simple charge diffusion model~\cite{chargediff} is implemented where the charge is allowed to diffuse uniformly, being reflected on the bottom of the epitaxial layer.
Charge is then accumulated by every pixel.
If the charge in a pixel is above a set threshold, a hit is created.
Recombination of charges is modelled by an attenuation length, however the tuning of the model to the data does not seem sensitive to the attenuation, so the length is set to infinity.

\section{Electromagnetic Shower Shapes}
\label{sec:showers}

\begin{figure}[h]
\centering 
\includegraphics[width=.9\textwidth]{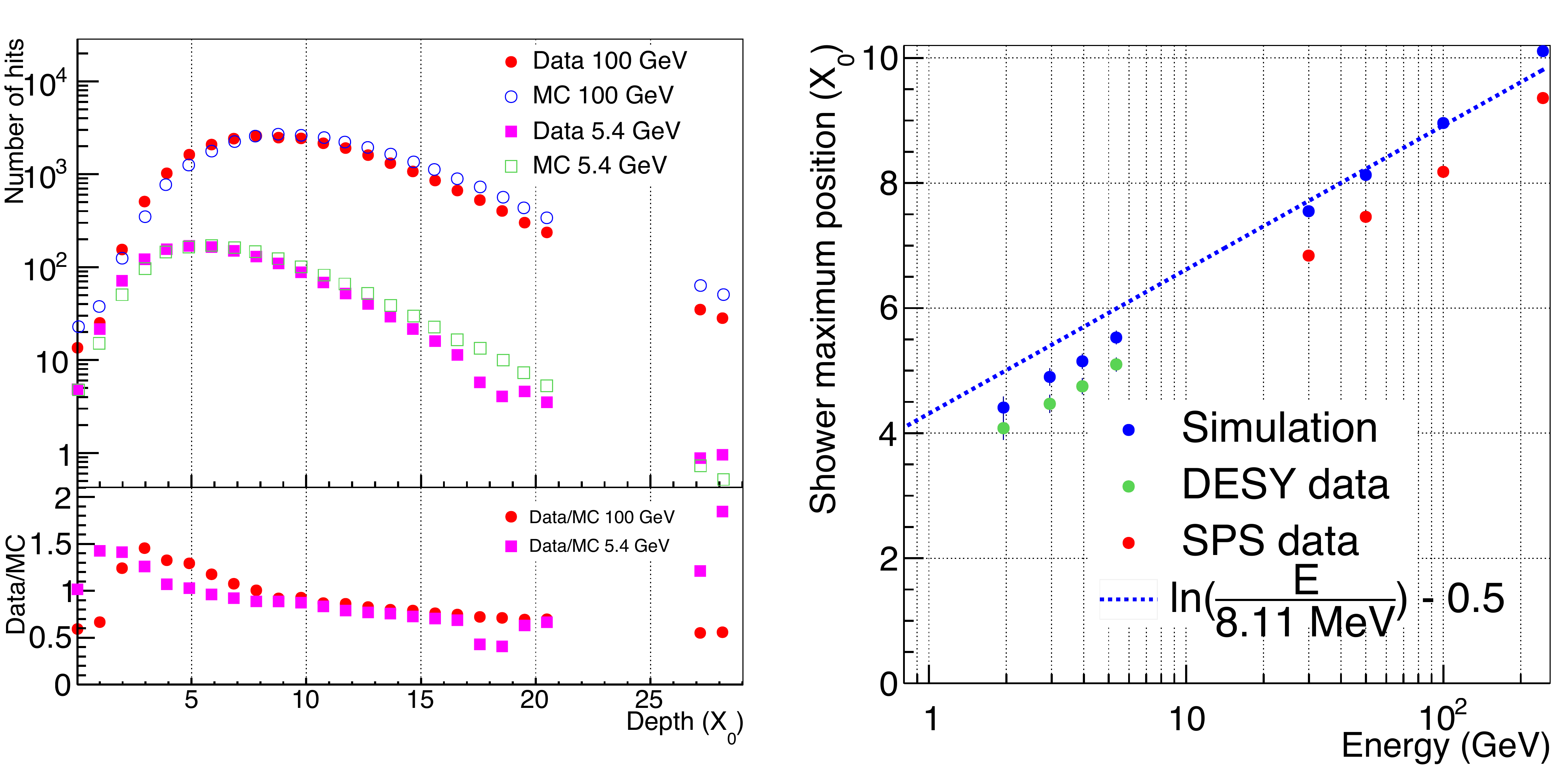}
\caption{\label{fig:longitudinalvsMC} (left) The longitudinal hit profile for 5.4\,GeV and 100\,GeV compared between data and MC simulations. (right) The shower maximum position as a function of beam energy for data and MC simulations. The theoretical expectation is shown as a dotted line.}
\end{figure}

\begin{figure}[h]
\centering 
\includegraphics[width=.9\textwidth]{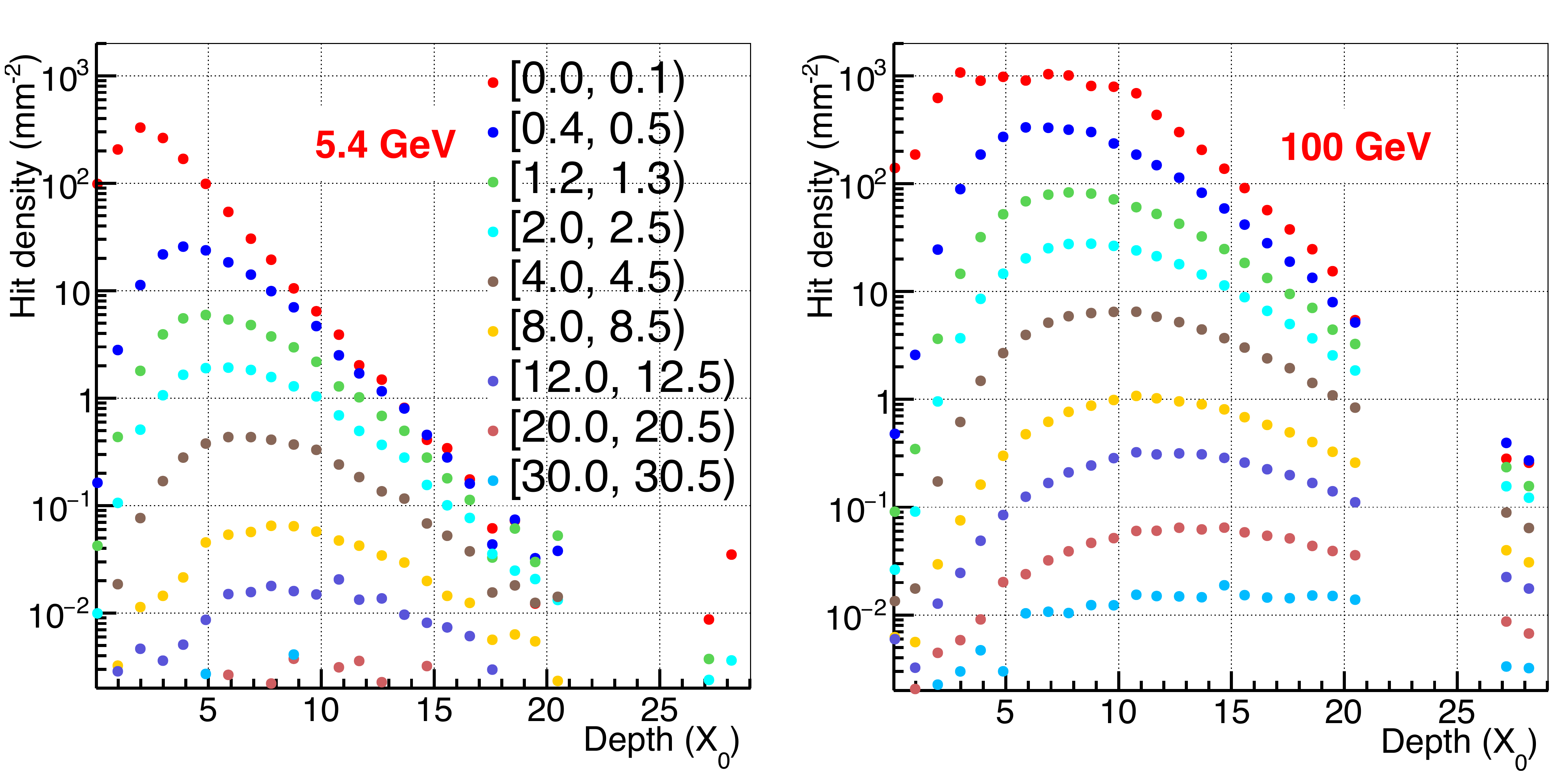}
\caption{\label{fig:longitudinalrings} The longitudinal hit density profile for 5.4\,GeV (left) and 100\,GeV (right) for different radial rings.}
\end{figure}

\begin{figure}[h]
\centering 
\includegraphics[width=.5\textwidth]{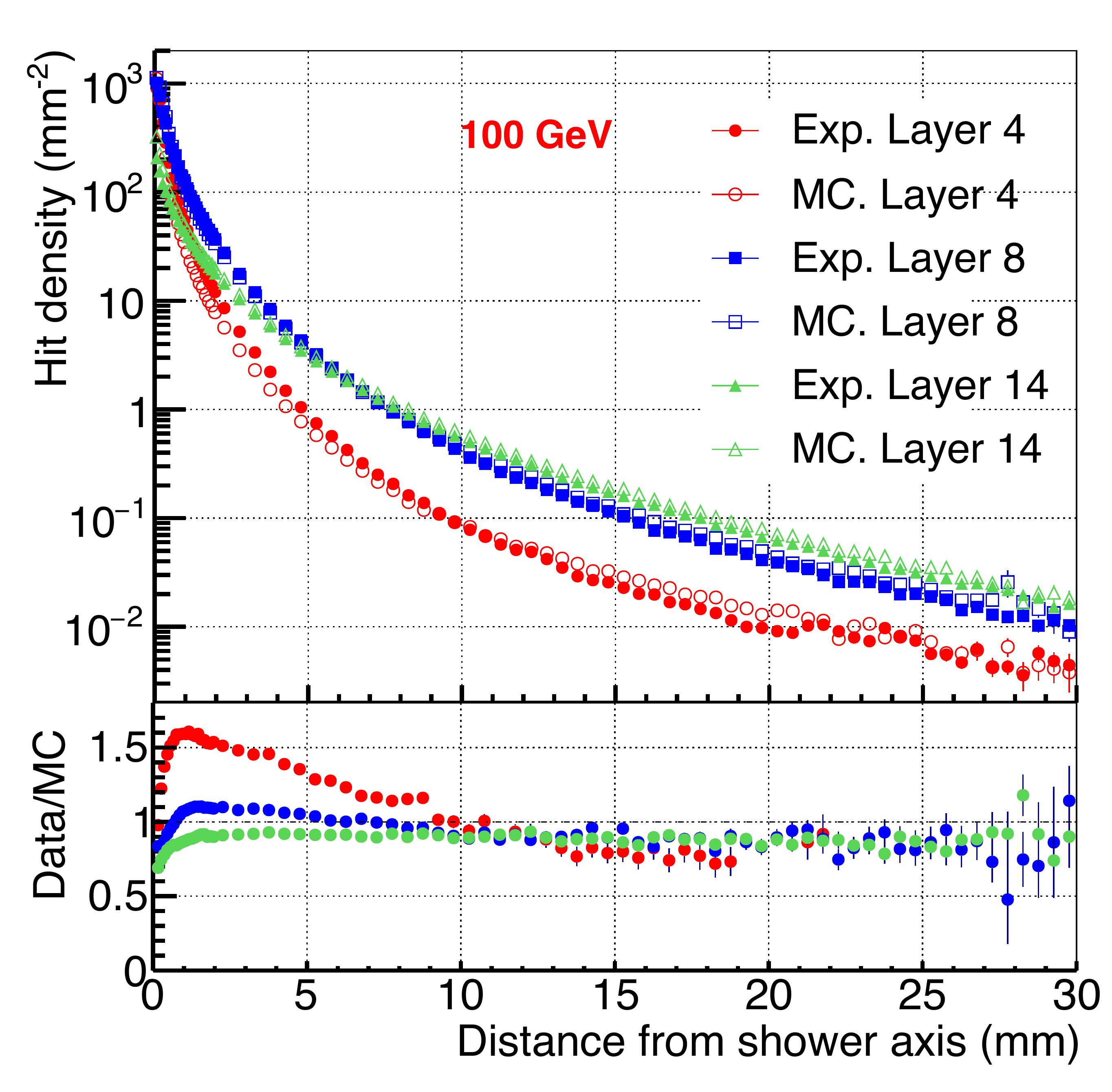}
\caption{\label{fig:radial} The radial hit density profile for 100\,GeV at a depth of layer 4, 8 and 14.}
\end{figure}

As mentioned above, all shower observables are reconstructed from the calibrated hit density.
For the longitudinal profiles the hit densities are integrated over the radial rings for each layer to get the total number of hits in that layer.
The longitudinal profiles for the different energies show that, as expected, the showers penetrate deeper into the material and the position where the maximum number of hits (shower maximum) occurs also moves deeper into the detector.
Figure~\ref{fig:longitudinalvsMC} (left) shows the longitudinal profile at 5.4\,GeV and 100\,GeV in data compared to MC simulations.
In the ratio one can see that in the first few interaction lengths the number of hits in the data is significantly larger than in the simulation. 
Figure~\ref{fig:longitudinalvsMC} (right) compares the shower maximum position between data and MC simulations.
The shower maximum is reached earlier in data, but both data and simulation do not match the theoretical expectation depicted by the dotted line.
Because of the high level of detail available in the prototype, the longitudinal profiles can also be made per radial ring from the shower centre.
These are shown in Fig~\ref{fig:longitudinalrings} for 5.4 GeV (left) and 100 GeV (right).
One can see that the maximum hit density moves deeper into the material for larger ring radii.
Additionally, for 100 GeV the effect of saturation is visible for the shower core (inner 0.1 mm).
The radial profiles show the average hit density as a function of ring radius per layer.
They show that the profiles broaden with depth, as expected.
The densities increase up to the shower maximum and then decay.
Figure~\ref{fig:radial} shows the hit density for layers 4, 8 and 14 for 100 GeV data and MC simulations.
The ratio clearly shows that the hit density in data in the shower core in the first few layers is larger than in the simulation.
Additionally, the MC profiles are narrower compared to the data.

These differences in data and MC simulations might be caused by an imperfect modelling of electromagnetic showers at this level of detail, or they might be an effect of the simplicity of the implemented charge diffusion model.
Within GEANT4 several options for electromagnetic physics are available~\cite{em}.
For the CALICE digital hadronic calorimeter~\cite{DHCAL} the EMY physics lists showed much better performance than the standard physics list.
EMY as well as EMZ offer the most precise calculation of standard and low energy models for precise electron, hadron and ion tracking.
For the FoCal simulations using the EMY and EMZ options results in only very small differences, 3-4\%, with the standard simulation.
While they occur in the same areas as the differences between data and simulation, the magnitude is not sufficient to compensate for the disagreement.
 
\section{Outlook}
\label{sec:outlook}
In the test beam data, because of the small spread in energy, electrons and pions can be very efficiently selected based on the total number of hits in the detector.
However, in a real experiment the energy of a particle is not known before it enters the calorimeter.
A high energy pion might create the same number of hits as a low energy electron/positron.
Therefor, ways to identify particles are being explored.
One way is to identify the starting depth of the shower.
Electrons and positrons on average start to shower sooner than pions.
The start of the shower, the interaction layer, is found based on the increase in the number of hits in subsequent layers.
The interaction for electrons/positrons occurs mostly before the 5th layer, that of pions is more uniformly distributed.
Another way that is being investigated is to fit the radial hit density in the inner 5 mm of the shower and compare the fit parameters between electrons and pions.
The distribution of the parameters for electrons is quite peaked, while for pions it is much more spread out.
This method showed some promising results~\cite{hongkai}.
A more recent study is attempting to model the 3D shower shape in an energy independent way, which should enable the identification of electromagnetic showers versus hadronic showers.

\section{Conclusion}
\label{sec:conclusion}
The FoCal digital electromagnetic prototype is a successful test of the MAPS technology and a proof of principle of digital pixel calorimetry.
It offers unprecedented detailed spatial distributions of particle showers, and electromagnetic showers have been studied.
This data offers a unique opportunity for testing MC simulation models in detail and the differences observed between FoCal data and simulations illustrates that improvements in electromagnetic models are still possible.
Effective particle identification methods for FoCal are being investigated.



\acknowledgments

We gratefully acknowledge the DESY and CERN accelerator staff for the reliable and efficient beam operations and the ALICE TC team for the general support for test beams at CERN.
This work was supported in part by the Chinese Scholarship Council and the Dutch research organisation NWO.




\begin{thebibliography}{99}

\bibitem{upgrade}
   T. Peitzmann, Upgrade of the ALICE Experiment, \emph{EPJ Web Conf} {\bf 71} (2014) 00106

\bibitem{CGC}
  L. McLerran and R. Venugopalan, Computing quark and gluon distribution functions for very large nuclei, \emph{Phys. Rev. D}, {\bf 49(5)} (1994) 2233-2241, Gluon distribution functions for very large nuclei at small transverse momentum, {\bf 49(7)} (1994) 3352-3355, Green's function in the color field of a large nucleus, {\bf 50(3)} (1994) 2225-2233

\bibitem{mimosa}
  R. Turchetta et al., A monolithic active pixel sensor for charged particle tracking and imaging using standard VLSI CMOS technology, \emph{Nucl. Instrum. Meth. A}, {\bf 458(3)} (2001) 677-689

\bibitem{prototype}
  A.P. de Haas, The FoCal prototype - an extremely fine-grained electromagnetic calorimeter using CMOS pixel sensors, arXiv:1708.05164

\bibitem{G4}
  S. Agostinelli et al., Geant4 - a simulation toolkit, \emph{Nucl. Inst. Meth. A}, {\bf 506(3)} (2003) 250-303

\bibitem{chargediff}
L. Maczewski et al., Study of cluster shapes in the Mimosa-5 pixel detector, \emph{Nucl. Inst. Meth. A}, {\bf 610} (2009) 640

\bibitem{em}
  http://geant4.web.cern.ch/geant4/collaboration/working\_groups/electromagnetic/physlist10.0.shtml

\bibitem{DHCAL}
  B. Freund et al., DHCAL with Minimal Absorber: Measurements with Positrons, arXiv:1603.01652

\bibitem{hongkai}
  H. Wang, Prototype Studies and Simulations for a Forward Si-W Calorimeter at the Large Hadron Collider, PhD thesis (2018)

  



\end{thebibliography}
\end{document}